\documentclass[%
 reprint,
 amsmath,amssymb,
 aps,
 onecolumn,
]{revtex4-2}

\usepackage{printlen}
\usepackage{color,soul}
 
\usepackage{xcolor}
\usepackage{graphicx}% Include figure files
\usepackage{dcolumn}% Align table columns on decimal point
\usepackage{bm}% bold math
\usepackage{hyperref}% add hypertext capabilities
\usepackage[mathlines]{lineno}% 
\usepackage{subfigure}

\newcommand{\matr}[1]{\mathbf{#1}}
\newcommand*{\affaddr}[1]{#1} 
\newcommand*{\affmark}[1][*]{\textsuperscript{#1}}
\usepackage{etoolbox}

\begin{document}

\preprint{APS/123-QED}

\date{\today}
\title{Growing polarisation around climate change on social media}

    \author{Max Falkenberg\affmark[1], Alessandro Galeazzi\affmark[2], Maddalena Torricelli\affmark[1], Niccolò Di Marco\affmark[3], Francesca Larosa\affmark[4,5], Madalina Sas\affmark[6], Amin Mekacher\affmark[1], Warren Pearce\affmark[7], Fabiana Zollo\affmark[2,8,*], Walter Quattrociocchi\affmark[9,*] and Andrea Baronchelli\affmark[1,10,*]
\begin{center}
\affaddr{\affmark[1]{ \textit{City University of London, Department of Mathematics, London EC1V 0HB, (UK)}}}\\
\affaddr{\affmark[2]{ \textit{Ca’ Foscari University of Venice — Department of Environmental Sciences, Informatics and Statistics, Via Torino 155, 30172 Venezia (IT)}}}\\
\affaddr{\affmark[3]{ \textit{University of Florence, Viale Morgagni, 40/44 - 50134 Florence (IT)}}}\\
\affaddr{\affmark[4]{ \textit{Institute for Sustainable Resources, University College London, London  WC1E 6BT, (UK)}}}\\
\affaddr{\affmark[5]{ \textit{Euro-Mediterranean Center on Climate Change (CMCC), Via delle Industrie, 13 - 30175 Venice (IT)}}}\\
\affaddr{\affmark[6]{ \textit{Centre for Complexity Science, Imperial College London, London SW7 2AZ, (UK)}}}\\
\affaddr{\affmark[7]{ \textit{iHuman, Department of Sociological Studies, University of Sheffield, Sheffield, S10 2TN, (UK)}}}\\
\affaddr{\affmark[8]{ \textit{The New Institute Centre for Environmental Humanities, Dorsoduro 3911, Venice (IT)}}}
\\
\affaddr{\affmark[9]{ \textit{Sapienza University of Rome — Department of Computer Science, Viale Regina Elena, 295 — 00161 Roma (IT)}}}
\\
\affaddr{\affmark[10]{ \textit{The Alan Turing Institute, British Library, London NW1 2DB, (UK)}}}\\
\affaddr{\affmark[*]{Corresponding authors: fabiana.zollo@unive.it, quattrociocchi@di.uniroma1.it, abaronchelli@turing.ac.uk}}
\end{center}
}
\begin{abstract}
\vspace{0.5cm}
\textbf{
Climate change and political polarisation are two of the 21st century's critical socio-political issues. Here, we investigate their intersection by studying the discussion around the UN Conference of The Parties on Climate Change (COP) using Twitter data from 2014 to 2021. First, we reveal a large increase in ideological polarisation during COP26, following low polarisation between COP20 and COP25. Second, we show that this increase is driven by growing right-wing activity, a 4-fold increase since COP21 relative to pro-climate groups. Finally, we identify a broad range of ``climate contrarian'' views during COP26, emphasising the theme of ``political hypocrisy'' as a topic of cross-ideological appeal; contrarian views and accusations of hypocrisy have become key themes in the Twitter climate discussion since 2019. With future climate action reliant on negotiations at COP27 and beyond, our results highlight the importance of monitoring polarisation, and its impacts, in the public climate discourse.
} 
\end{abstract}
\maketitle
\vspace{0.2cm}

Social media platforms, {like Twitter}, provide important locations for the everyday discussion and debate of climate change \cite{pearce2019social}. The nature of this role is highly contested, with some pointing to its democratising potential, while others argue that social media is accelerating political polarisation \cite{schafer2019social}. \textcolor{black}{Monitoring polarisation is important given that a highly polarised environment has the potential to} drive antagonism between ideological groups, generate political deadlock, and threaten pluralist democracies \cite{mouffe2014way}. \textcolor{black}{Thus, the study of online polarisation has gained momentum in recent years} \cite{barbera2015birds,flamino2021shifting,bovet2019influence}.

In this paper, we analyse tweets related to the Conference of the Parties (COP) in order to clarify the nature of polarisation in political debates on climate change. \textcolor{black}{Specifically, we are interested in how the climate discussion is structured on Twitter in terms of the plurality of views and the interaction patterns amongst ideologically opposed groups. We find that a prominent opposition to the dominant pro-climate discourse has established itself since late 2019, resulting in a highly polarised online climate debate.} 

Twitter is the ideal platform for studying climate communication as it is widely used by politicians and journalists \cite{schafer2019social}{, has broad social and cultural influence} \cite{burgess2020twitter}, 
{and because of the rich structural data it captures.} {Of course, Twitter is not directly analogous to public opinion, and our results likely derive from a combination of the platform's well-documented tendency to foster polarisation and the broader contexts for climate politics} \cite{marres2017digital,williams2015network,urman2020context}. However, many studies highlight the importance of Twitter, \textcolor{black}{and social media in general}, as a critical tool for studying climate communication \cite{oneill2015dominant,pearce2014climate,walter2019scientific,cody2015climate,chen2021polarization,fownes2018twitter,pearce2019social}, {political polarisation} \cite{bhadani2022political,bovet2019influence} {, and misinformation} \cite{pennycook2021shifting}.  \textcolor{black}{Beyond social media, a broad literature considers the polarisation and politicisation of climate issues using other computational techniques and more traditional approaches} \cite{lucas2018concerning,mccright2011politicization,drummond2017individuals,farrell2016ideological,chinn2020politicization}. \textcolor{black}{Here, we extend this literature by exploiting tools from the growing field of ``infodemics''} \cite{tucker2018social,del2016spreading,cinelli2020covid,cinelli2021echo,bessi2015science,flamino2021shifting,gallotti2020assessing,zarocostas2020fight,bail2018exposure}.

The motivation for our focus on COP is threefold. {Firstly, the COP discussion can be characterised as a discrete, regularly repeated online event which lends itself to a quantitative, multi-year analysis of climate polarisation (a key gap in the literature). Secondly, by focusing on a specific event we ensure that tweet content is thematically focused (in our case on climate politics) and that the network of interactions is sufficiently connected to allow robust network analysis (which is not always possible with sampled datasets} \cite{wang2015should}). This event-focus is a common feature of previous climate communication studies on Twitter, (e.g., on the IPCC report \cite{pearce2014climate} {or the Finnish elections} \cite{chen2021polarization}). {For a review of the benefits of studying specific events or controversies, see} \cite{venturini2021controversy}.
{Finally, COP is the pre-eminent international forum for climate diplomacy, directing considerable public attention towards climate change} \cite{boykoff2021review,schmidt2013media,hopke2018visualizing}. {This makes COP the ideal target for studying the intersection between climate change and political polarisation.} 

{Here, we first highlight the significance of COP21 and COP26 relative to other COPs.}
Second, we derive a spectrum of ``climate ideologies'' (defined by constructing a synthetic distribution of opinions based on similarities in user-user interactions), which reveals two prominent groups: an ideological minority and majority.
We reveal that polarisation {(measured as the bimodality of the ideology distribution)} is low pre-COP25, before a large increase in COP26 \textcolor{black}{(with supplementary data suggesting that the increase in polarisation likely started in 2019 around the global climate strikes)}. 
Third, we emphasise the political dimension of COP, revealing broad international engagement from elected politicians, and highlighting the political parties who oppose urgent climate action. Fourth, we investigate discussion topics during COP26 and highlight the overlap between minority rhetoric and established ``climate contrarian'' views \cite{coan2021computer}. Notably, the issue of ``political hypocrisy'' is identified as salient issue of cross-ideological appeal. \textcolor{black}{Finally, we supplement our analysis with Twitter data on climate scepticism and climate change and show that our COP analysis is broadly representative of the wider climate discussion on Twitter. }

\section*{Results}

We start by highlighting the significance of COP21 and COP26 (Figure~1). Panel (a) shows the number of posts for Twitter from 2014 to 2021. The inset shows general online engagement with COP measured using Google Trends, revealing that Twitter engagement closely reflects wider online attention. Within our study period, COP21 and COP26 are of particular significance, with the Paris Agreement signed at COP21, and the Glasgow Climate Pact agreed at COP26. Consequently, content creation and engagement {(i.e., retweet count)} are larger for COP21 and COP26 than in the intermediate years. 
Our data shows the influence of local engagement (see inset) where overall Google Trends scores are presented alongside country specific scores for France (the host of COP21) and Great Britain (the host of COP26). \textcolor{black}{Supplementary information (SI) sections 1A and 1B show a similar analysis for Youtube and Reddit, where activity is significantly lower than on Twitter.}

\begin{figure}[h!]
    \centering
    \includegraphics[width=0.98 \linewidth]{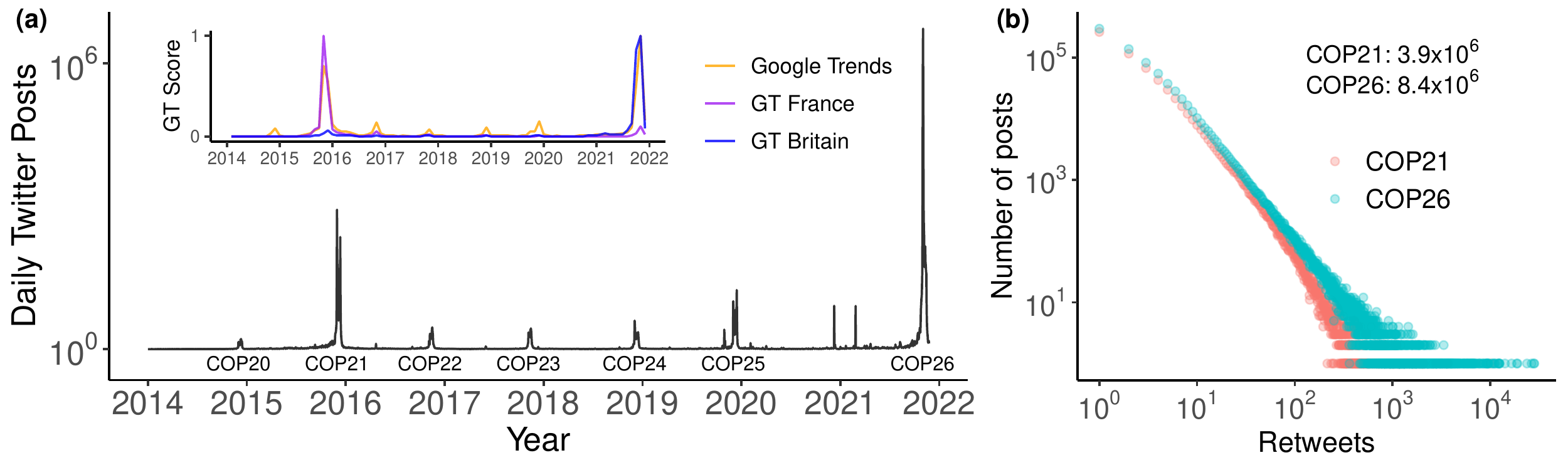}
    \caption{\textbf{Content creation and user retweet distributions on Twitter from COP20 to COP26}. (a) Total number of Twitter posts created each day using the term ``COP2x''. Inset: Google trends (GT) popularity score for ``COP2x'', with country specific scores showing the local enhancement of public engagement. (b) The retweet distributions for COP21 and COP26; inset: total number of retweets. Extended time period and other COPs shown in Supplementary Figures 1 and 2. }
    \label{fig:engagement}
\end{figure}

\subsection*{Ideological polarisation during COP}
\textcolor{black}{To assess the emergence of a broad climate contrarian community,} we now analyse the evolving nature of ideological polarisation
between COP20 and COP26. {Polarisation is most often quantified in terms of the modality of a distribution of surveyed opinions \cite{bramson2017understanding} (the definition we choose to use here; see discussion in Methods)}, {although other valid definitions exist}. \textcolor{black}{However, on Twitter true opinion data is unavailable, so instead we infer a synthetic opinion distribution from retweet data as a proxy (see Methods).} In subsequent sections, \textcolor{black}{we validate this proxy-method by showing how opposite ends of the ideological spectrum correspond to distinct views on climate change.}

We start by assuming that the climate ideology of an individual, $i$, can be expressed as a single number, $x_i$ \cite{cinelli2021echo}. Then, polarisation refers to the properties of the probability distribution, $\mathcal{P}(x)$, of ideological scores across a population. 
The ideological spectrum is extracted from the Twitter retweet network using the ``latent ideology'' method \cite{barbera2015tweeting,barbera2015birds,flamino2021shifting}. {Loosely speaking, the method produces an ordering of users and influencers where accounts with similar retweet interactions are close to each other in the ordering, resulting in similar scores (see Methods).} 
We specify that the majority group map to $-1$ and the minority to $+1$. {Any account with an ideology score less than (more than) zero is part of the majority (minority).} 

We calculate the latent ideology for COP21 and COP26 (Figure~2), where influencers are selected as the top 300 most retweeted accounts, excluding a small number ($3\%$) which conflate the results (see Methods). \textcolor{black}{Influencer demographics and labelling are discussed in SI section 1D.}

\begin{figure}[h!]
  \centering
  \includegraphics[width=.98\linewidth]{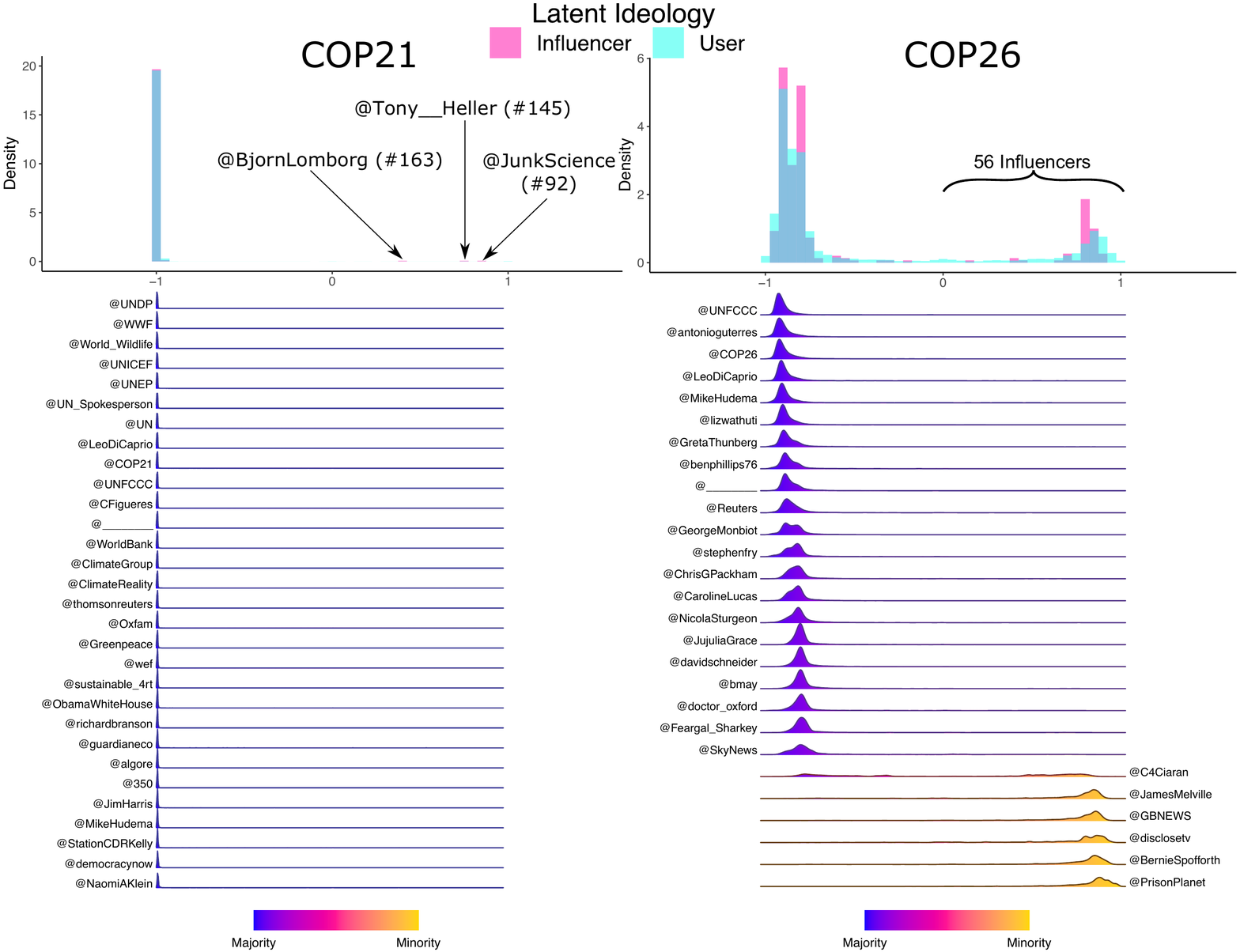}
    \caption{\textbf{The ideological spectrum for COP21 and COP26.} Top: A histogram of the influencer and user ideology scores for COP21 and COP26. The ideological minority map to $+1$, whereas the majority group map to $-1$. Bottom: The 30 most retweeted influencers and accompanying user ideology distributions. Influencers who are primarily retweeted by the ideological minority are written on the right, and influencers primarily retweeted by the ideological majority are on the left. Alongside each influencer we show the distribution of user ideologies who retweeted that influencer. For COP21, no members of the ideological minority are found amongst top influencers, in contrast to COP26 where we observe ideological polarisation. Note, @C4Ciaran appears in the minority but has cross-ideological appeal due to tweets referencing diesel car emissions, see section on ``Political hypocrisy \& the ideological divide'' below. Expanded figure with all 300 influencers is available \href{https://osf.io/nu75j/?view_only=53e03939cd824bc680e83b7c64c80b27}{here}. Active users with fewer than 30,000 followers indicated with @\_\_\_\_\_\_\_. Other COPs see Supplementary Figures 11--15.}
    \label{fig:ideology1}
\end{figure}

The latent ideology shows unimodal user ideology for COP21 whereas the COP26 user ideology is multimodal, 
as confirmed by Hartigan's diptest (see Methods): the \textcolor{black}{bimodality statistic}, $D$, increases from COP21 to COP26
(COP21: $D = 0.0023$, $95\%$ CI: [$0.0020,0.0026$], $p = 0.003$; COP26: $D = 0.049$, $95\%$ CI: [$0.048,0.050$], $p < 2.2 \times 10^{-16}$). {Despite the special significance of COP21, similarly low polarisation (i.e., unimodal ideologies) is found for all COPs prior to COP26 (see Extended Data 1).} 

\textcolor{black}{For both COP21 and COP26, influencers split into majority and minority actors. The majority are largely pro-climate accounts. Focusing on the minority gives some indication of the ideological divide present in these datasets.} The COP21 minority has three influencers: @BjornLomborg, @Tony\_\_Heller, and @JunkScience. These individuals are climate-focused and self-identify as outside the climate mainstream: @JunkScience quotes a Nature Climate Change article referring to him as ``the most influential climate science contrarian'' \cite{farrell2019evidence}, @BjornLomborg references his book ``False Alarm: How Climate Change Panic Costs Us Trillions'', and @Tony\_\_Heller links to his climate-critical blog ``realclimatescience.com''. 

For COP26 we find 56 minority influencers. Of these, 6 have a clear climate focus. The remainder include media organisations and journalists (e.g., @newsmax, @nypost, @GBNEWS, @PrisonPlanet, @bennyjohnson), politicians (@SteveBakerHW, @laurenboebert), and accounts campaigning against Covid-19 restrictions (@BernieSpofforth, @JamesMelville). This last group may not have strong view on climate, however, their presence in the minority remains important given how similarities in user-content interactions are used by recommendation systems~\cite{goel2015follow}. 

Qualitatively, the increase in polarisation is robust to variable influencer number, {different influencer definitions, different data-collection time windows, and the removal of tweets related to Covid-19} (see Supplementary Figures 3--7). Further analysis also suggests that bot activity and deleted content do not conflate the observed increase in polarisation (see SI sections 2C and 2D). \textcolor{black}{Our analysis shows that around 30\% of climate sceptic accounts from 2015 are no longer active on Twitter. However, deletion rates would need to exceed 80\% in order to explain the observed increase in polarisation.}

One important question is whether growing polarisation is a consequence of shifting views {(i.e., individuals moving from a majority to minority ideological position), or changes in minority activity {(i.e., users with pre-existing climate sceptic views expressing those views more prominently on Twitter)}. We assess this by recomputing the ideological spectrum using an equal number of minority and majority influencers who appear in both the COP21 and COP26 datasets} (Figure~3). This shows that minority influencers from COP21 remain in the minority for COP26, and majority influencers remain in the majority; \textcolor{black}{over half of minority influencers selected using this method are climate-focused accounts. However, for the standard retweet-based COP26 minority (Figure~2), only 11\% are climate-focused. This demonstrates that the promotion of climate contrarian views is shifting away from climate-focused accounts towards a broader set of non-specialised influencers.} {The observed increase in polarisation is likely due to users with existing minority views expressing those views more prominently on Twitter, although determining this precisely is difficult.}  

\begin{figure}[h!]
  \centering
  \includegraphics[width=.98\linewidth]{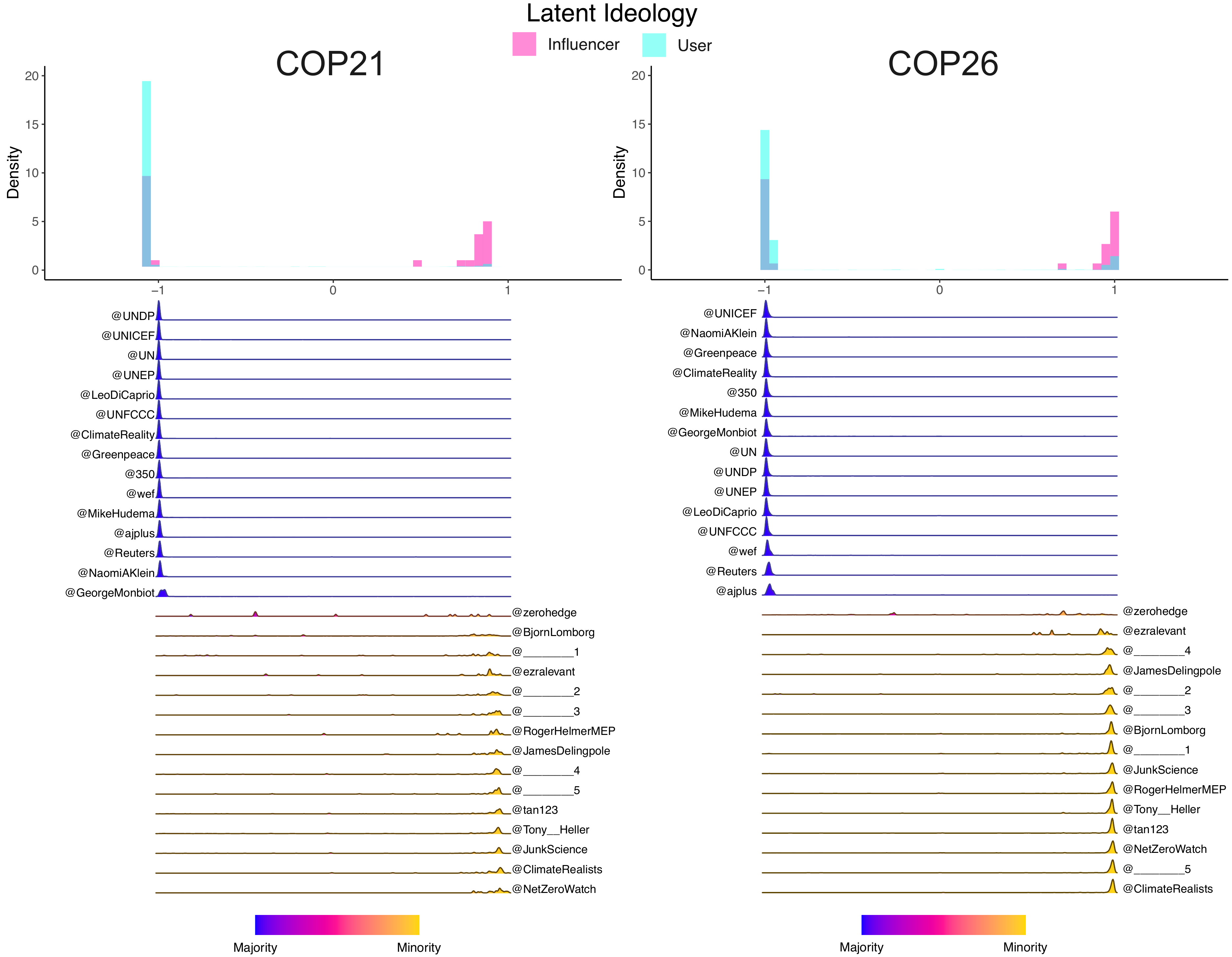}
    \caption{\textbf{The ideological spectrum for COP21 and COP26 recomputed using an equal number of minority and majority influencers.} Majority (minority) influencers are written on the left (right) of each panel. Influencers selected must appear in both the COP21 and COP26 datasets. Influencer polarisation is similar between COP21 and COP26, but user polarisation (i.e., distribution bimodality) increases significantly. This reflects a large increase in user engagement with the ideological minority during COP26 (i.e., minority influencers are attracting a disproportionately large fraction of retweets in COP26 relative to COP21). More detail is provided in SI section 1G. Active users with fewer than 30,000 followers indicated with @\_\_\_\_\_\_\_, excluding elected politicians (@RogerHelmerMEP).}
    \label{fig:ideology2}
\end{figure}

\subsection*{The political dimension of COP26}

\begin{figure}[h!]
    \centering
    \includegraphics[width = \linewidth]{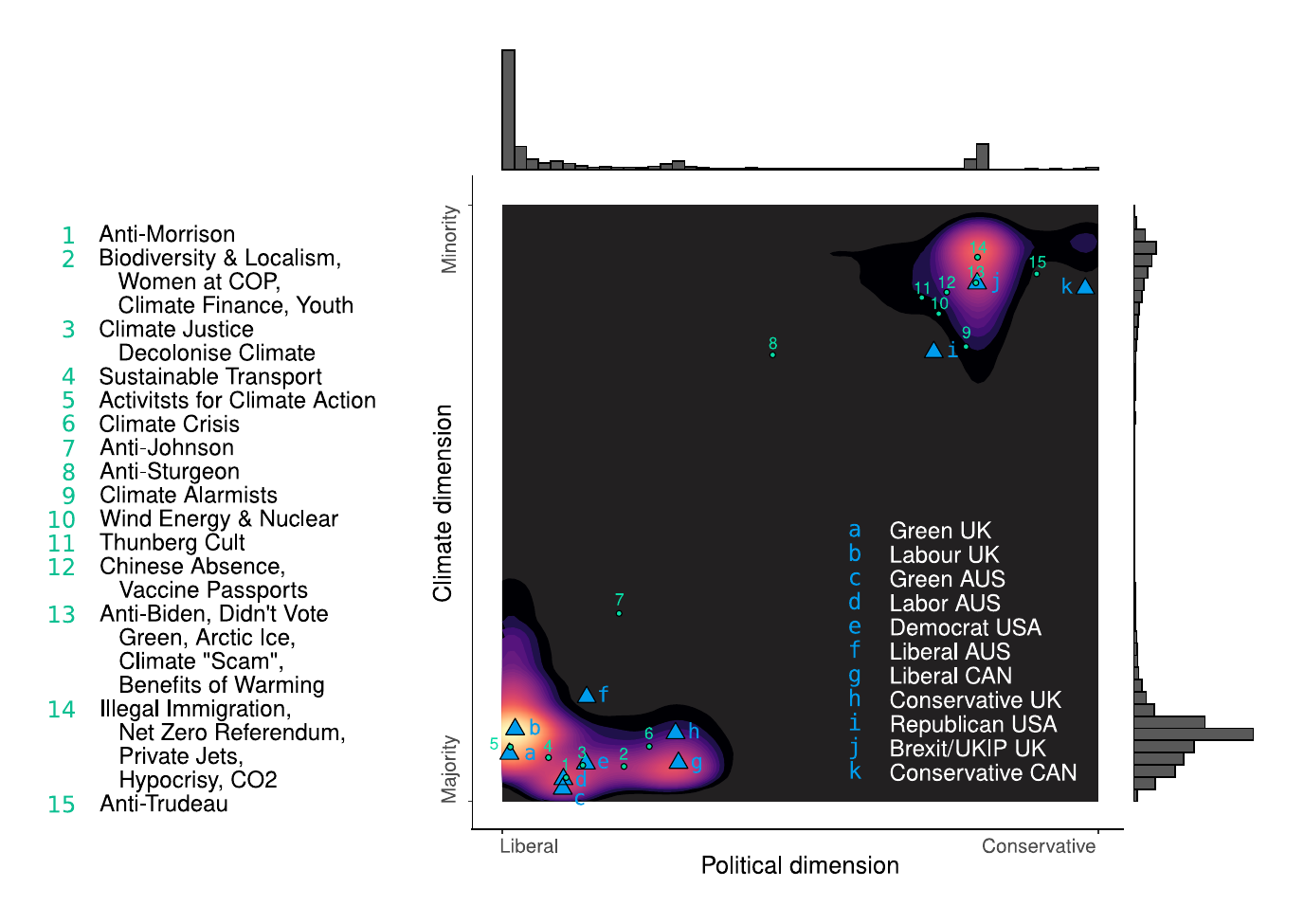}
    \caption{\textbf{A two-dimensional representation of the latent ideology, split according to political and non-political influencers.} Triangular points label the median ideological position of accounts affiliated with specific political parties. Circular points indicate the median position of users who tweeted a particular topic, as derived using BERTopic. In Fig.~2, the latent ideology is calculated using the top 300 most retweeted accounts. Here, we calculate the latent ideology twice using (1) the top 300 most retweeted accounts affiliated with individual elected politicians (horizontal axis; political dimension), and (2) using the top 300 most retweeted accounts excluding politicians (vertical axis; climate dimension). The non-political axis can be thought of as the general climate dimension, whereas the political axis can be thought of as capturing the specific political groupings of the COP discussion. Some topics are merged into a single point for visual clarity. }
    \label{fig:politics}
\end{figure}

\textcolor{black}{To better understand how ideological views on climate change are associated with political leanings,} we now highlight the role of elected politicians in the COP26 dataset (see Methods). We do this by recomputing the latent ideology twice: first, using exclusively politicians as influencers, and second, excluding politicians,  generating a two-dimensional spectrum (Figure~4). {Political engagement between COP20 and COP25 is discussed in SI section 1E; for COP21 we find only one minority politician (Roger Helmer, former UKIP MEP).}

Marking the median position of select Anglophone political parties shows how the majority and minority, which appear homogeneous along the climate axis, split into groups with more geographical and political nuance. 
{In the minority, we find a large block dominated by the US Republicans and former UK Brexit/UKIP politicians, alongside a smaller block corresponding to the Canadian Conservative party. Related tweet extracts include calls for a ``Net Zero Referendum'' (Nigel Farage), claims that COP ``has absolutely zero credibility'' (Lauren Boebert), and statements that ``everything you've being [\textit{sic}] told by climate alarmists is a lie'' (Maxime Bernier).}

{In the majority we find most other mainstream political parties. It is perhaps surprising that some parties criticised as weak on climate action appear in the majority, notably the Australian Liberals} \cite{crowley2021fighting}. {However, this reflects pro-climate rhetoric by Scott Morrison which attracted majority retweets (e.g., ``pleased to agree a new low emissions tech partnership''), but also criticism (``\#ScottyFromMarketing'').}

One apparent oddity is that left-leaning political groups (e.g., UK Labour and the Greens) appear ideological closer to the minority on the climate-axis than more conservative parties. \textcolor{black}{Analysis below suggests that this is due to cross-ideological accusations of political hypocrisy.}

\subsection*{Topics of discussion}

{Topics in the COP26 discussion can be extracted using BERT topic modelling} \cite{grootendorst2022bertopic}{ (see Methods) and placed on the ideological spectrum (Figure~4). See SI section 1I for COP21 results.} 

\subsubsection*{Majority topics}

{Majority topics have a clear climate focus, making explicit reference to specific COP themes, including ``women's day'', ``transport day'', and ``climate finance''. Beyond these, there are topics related to climate activism with a specific emphasis on youth protests, indigenous groups, the need for ``climate justice'', and the ``decolonisation'' of climate change.} 

{Potentially the most striking rift between users in the majority relates to whether they are COP-supportive or not. Many pro-climate accounts are  critical of the COP process, describing it as ineffective and accusing it of ``greenwashing''.
This theme is a clear shift from COP21 where only select influencers were COP-critical (7\% of labelled COP21 influencers, see SI section 1D), most notably George Monbiot and Naomi Klein.
However, criticism of the COP process has grown significantly (35\% of labelled COP26 influencers).} 

\subsubsection*{Minority topics}

The COP26 minority discuss a broad range of climate related topics. Cross-referencing these with a taxonomy of ``climate contrarian'' claims \cite{coan2021computer}  shows that the COP26 minority promote, and engage with, all five of the leading contrarian claim types (Table 1). 

{Other topics not specific to climate include tweets critical of particular politicians, most notably Joe Biden (referred to as ``sleepy Joe''), Boris Johnson (for promoting green policies), and Justin Trudeau (for allegedly destroying the Canadian Oil/Gas industry). Finally, there are topics of wider relevance to the political right, particularly Covid-19 (the ``plandemic''), vaccines (``\#NoVaccinePassports''), and illegal immigration (``[stop] illegal economic migrants'').}

\begin{table}[h!]
\begin{tabular}{|l|l|l|}
\hline
\textbf{Claim type}                                                                                    & \textbf{Topics} & \textbf{Representative tweet extracts}                                                                                                                                                                                                                                                                                                                                             \\ \hline
(1) - Global warming isn't happening                                                                   & \begin{tabular}[c]{@{}l@{}} 9, 13, 14     \end{tabular}        & \begin{tabular}[c]{@{}l@{}}``ALL the current data say the opposite. \\ Arctic Ice Extent at a 6 year high, volume up 17\% on last year \\ when all models predict most warming will occur at The Poles. \\ You're talking shite" \\  ``@COP26 You have been lying to the public and mocking them \\ for decades with your climate scam"\end{tabular}                             \\ \hline
\begin{tabular}[c]{@{}l@{}}(2) - Human greenhouse gases are \\ not causing global warming\end{tabular} & 13, 14              & \begin{tabular}[c]{@{}l@{}}
``What \% of Atmosphere is made up of CO2??\\ 
A:  0.04\% \\  
Of that 0.04\%, humans create 3\%, or 0.0012\% \\ 
= FCUK ALL!!"\end{tabular}       \\ \hline
(3) - Climate impacts are not bad                                                                      & 9, 13             & \begin{tabular}[c]{@{}l@{}}``@COP26 What `climate change'? \\ Thriving polar bears. Record coral cover. \\ Stable ice-sheets. Bumper snow. \\ Greening planet. Reduced wildfires. \\ Increased Pac island land. Fewer hurricanes.  \\ Antarctica record cold winter. \\ CO2 causation not proven. CO2 beneficial."\end{tabular}                                                    \\ \hline
(4) - Climate solutions won't work                                                                     &             \begin{tabular}[c]{@{}l@{}}
10, 12, 13, \\ 14 
\end{tabular}& \begin{tabular}[c]{@{}l@{}}``We didn’t vote for this impoverishing green socialist nonsense."\\ ``China is not going to COP26. So what’s the point?"\end{tabular}                                                                                                                                                                                                               \\ \hline
\begin{tabular}[c]{@{}l@{}}(5) - Climate movement/science \\ is unreliable\end{tabular}                & \begin{tabular}[c]{@{}l@{}}
9, 11, 13,\\  14, 15   
\end{tabular}          & \begin{tabular}[c]{@{}l@{}}``It’s amazing - and desperate IMO - that alarmists are STILL hanging\\ on to the presumption that the Maldives are about to go underwater any\\  time soon, even after 30 odd years of failed claims that it is imminent!" \\ ``More blah blah blah from the Thunberg cult, none of whom have \\ reduced their carbon footprint 1 jot"\end{tabular} \\ \hline
\end{tabular}
\caption{\textbf{{Common claims made by groups who oppose climate action and examples from the COP26 minority.}} {The left column lists the five leading claim types made by ``climate contrarians" according to} \cite{coan2021computer}.
{For each claim we list related topic numbers from the ideological minority} (Figure~4) {extracted from the COP26 dataset. Each claim is accompanied by representative tweet extracts. These are detected automatically by using the BERTopic ``representative document'' function.}}
\label{tab:climate_claims}
\end{table}

\subsubsection*{{Political hypocrisy \& the ideological divide}}

\textcolor{black}{Understanding content that bridges the ideological divide is important for assessing which topics may act as a gateway into the ideological minority, particularly since Twitter recommends content based on similarites in user-content interactions between accounts} \cite{goel2015follow}.

{To assess this, we rank tweets according to the number of cross-ideological retweets, i.e., majority author but minority retweeter, or vice-versa. This reveals the theme of ``political hypocrisy'' which includes references to the use of private jets and diesel cars, the continued use and development of fossil-fuels, and the dumping of raw sewage.} Half of all majority tweets referencing hypocrisy have been posted since December 2020 (SI section 1K; {Extended Data 2).}

\subsection*{News media reliability}

\textcolor{black}{Given the distinct topics discussed by the majority and minority, we may expect that these ideological groups reference different news media outlets.} We show this using heatmaps of ideology against independent news media reliability scores (Figure~5; see Methods). This reveals that the ideological majority preferentially reference news domains with high trust scores, whereas the minority often reference domains with low scores. {This result is robust if we use country-specific NewsGuard scores (SI section 1F).} \textcolor{black}{In SI section 1J, we show the formation of ideological ``echo chambers'' during COP, a common feature of polarised communities on social media} \cite{williams2015network,cinelli2021echo}.

\begin{figure}[h!]
    \centering
    \includegraphics[width =0.9 \linewidth]{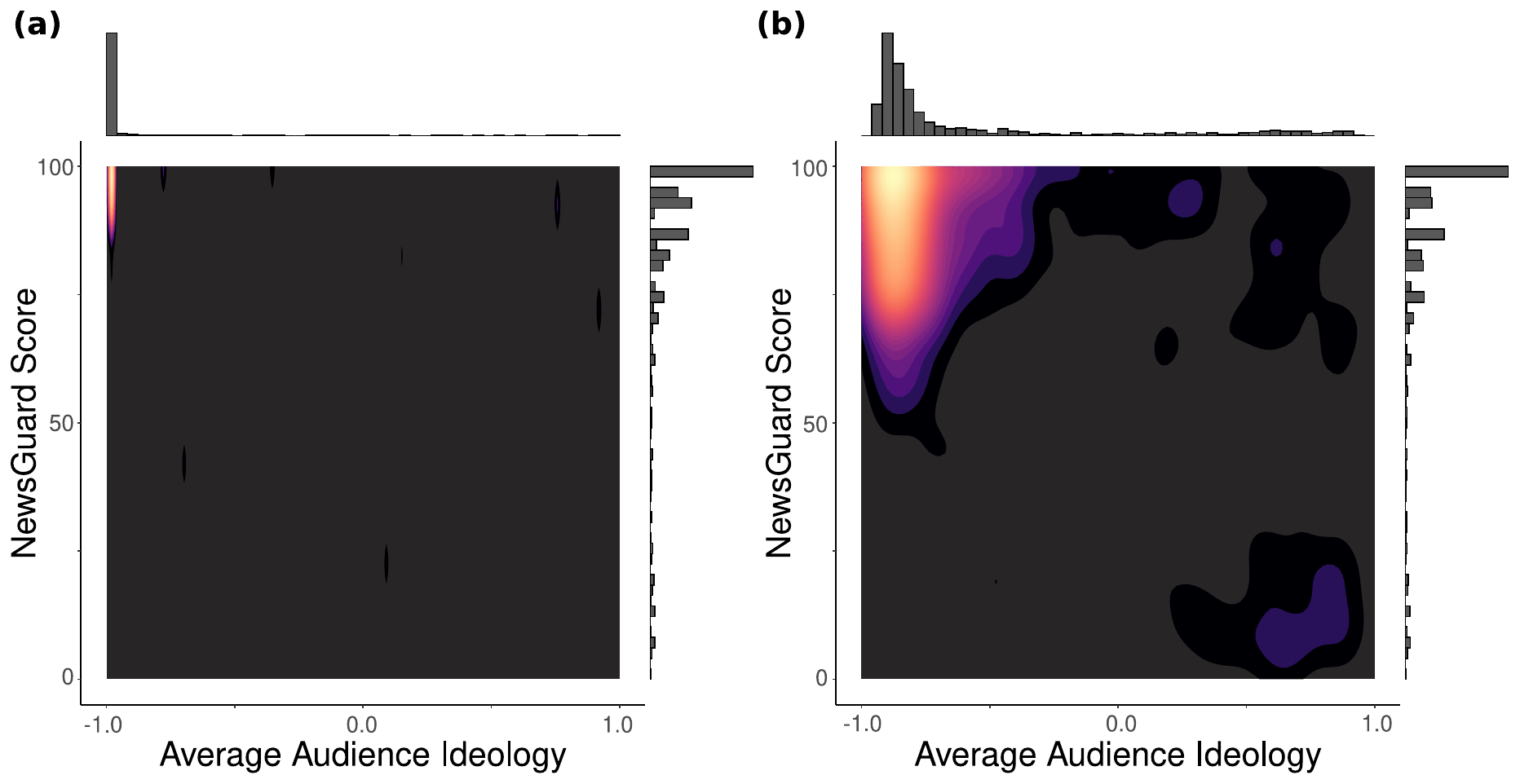}
    \caption{\textbf{The correlation of climate ideology with distinct media outlets.} Heatmaps showing the density of news media trust scores provided by NewsGuard, against the average ideological score of each media outlet's Twitter audience for (a) COP21 and (b) COP26. Visualisations of the COP21 and COP26 network community structure are shown in SI section 1J, alongside heatmaps illustrating the ``echo chamber'' effect, see \cite{cinelli2021echo}.}
    \label{fig:communities}
\end{figure}

\subsection*{{The wider climate discussion on Twitter}}

We now show that the COP discussion is broadly representative of the wider climate discussion on Twitter. {This is important since keyword-based data collection using the search term ``COP2x'' may fail to capture certain climate-related communities.}

\textcolor{black}{First, we cross-reference our COP dataset with two supplementary datasets: 1.3 million tweets using terms associated with climate scepticism, and all original tweets since 2012 using the term ``climate change'' (SI section 1K).} This reveals that (1) the activity of the COP26 minority is highly correlated to the activity of the broader climate sceptic community on Twitter {(Extended Data 3)}, (2) the COP26 minority started to engage with climate issues much more recently than the COP26 majority {(Extended Data 4)}, and (3) climate sceptic activity was very low, but present, before 2019. {Note that there is no evidence of a change in the interaction rate between pro-climate and climate contrarian groups (SI section 1K).}

{The expression of} climate scepticism saw significant growth from 2019 onwards, peaking during the global climate strikes in September 2019, and the Australian bushfires in January 2020. This growth does not appear to have translated into significant engagement from sceptics during COP25, most likely due to its lesser importance (Figure~1; major new agreements were not negotiated at COP25).  

{Increased sceptic activity does not necessarily imply an increase in the number of Twitter users with climate sceptic views, but more likely an increase in users expressing those views (which is in itself important)}. Possible drivers of this growth include (1) the issue of political hypocrisy (see above), specifically following the approval of a new Canadian oil pipeline in June 2019, (2) as a backlash to the direct impact of the global climate strikes (minority content is particularly critical of Greta Thunberg and Extinction Rebellion), and (3) the belief that the climate movement is unreliable, with minority users blaming the Australian bushfires on arson, not climate change.  

\section*{Discussion \& Conclusion}

We have investigated ideological polarisation around climate change by analysing the discussion around COP on Twitter. Our results show that ideological polarisation, {measured in terms of bimodality}, was low and largely flat between COP20 and COP25, before a significant increase in COP26, driven by growing right-wing {activity}. 

\textcolor{black}{Cross-referencing the COP dataset with additional data on climate scepticism highlights 2019 as a key year {where the expression of} climate scepticism grew on Twitter. Our data points towards the role of political hypocrisy, and a potential backlash to direct action from climate activists (see }\cite{sarewitz2010curing,merkley2021party,vihma2021climate} \textcolor{black}{for related discussions) as potential factors in this growth.} 

{The opposition to climate action is a known feature of populist politics} \cite{huber2021populism}, {largely due to the association of climate change with issues of institutional trust, and populist attitudes towards science.} {This trend has likely been catalysed by anti-science sentiments during the Covid-19 pandemic} \cite{prasad2022anti}. However, surveys suggest that right-wing views on climate are more subject to change than left-wing views \cite{jenkins2020partisan}. Consequently, there is reason to believe that growing right-wing opposition to climate action may be reversible. 

\textcolor{black}{It is perhaps surprising that the events with the greatest increase in climate scepticism on Twitter took place since 2019} and not earlier, particularly given Trump's election \cite{waller2021quantifying,flamino2021shifting}and Brexit \cite{del2017mapping}.
{Climate issues were not a central feature of the 2016 Brexit debate. Yet, many members of the COP26 minority were prominent Brexit campaigners. This shift may be a sign that these politicians see opposition to climate action as a topic with growing popular appeal; note for example Nigel Farage's ``Net Zero Referendum'' campaign.} 

Given that rapid and effective climate action is dependent on broad international consensus and collaboration, \textcolor{black}{the growth in polarisation may risk political deadlock if it fuels antagonism to climate action} \cite{mouffe2014way}. Policy makers should consider how actionable factors may be driving this polarisation; perceptions of political hypocrisy may be critical in this regard. Our analysis suggests that these perceptions are worsening, not improving. \textcolor{black}{Similar concerns regarding hypocrisy discourse around climate change have been raised previously} \cite{gunster2018don,xia2021spread}. {For instance, researchers have shown that tweets referencing climate hypocrisy are associated with an increase in tweet virality} \cite{xia2021spread}. 

Our analysis focuses on Twitter since this is where the COP discussion is most active and where we find influencers from across the political spectrum. {Data was acquired using keyword search (``COP2x'') which ensures that tweets are thematically focused and data can be feasibly acquired from the Twitter API. In principle, this approach may fail to capture the full climate conversation on Twitter, but supplementary data shows that our results are broadly representative of the wider climate discussion.}
Future work should acquire larger datasets {with a broader focus} {(perhaps using a random tweet sample if sufficient data can be acquired, although this approach is problematic for structural analysis} \cite{wang2015should}), and could consider a wider range of platforms \cite{sipka2021comparing}. Our data suggests that the COP discussion is not particularly active on Youtube or Reddit, although this may not be the case for other climate events. 

Given \textcolor{black}{significant engagement with climate politics during COP26 from groups and politicians opposed to climate action}, future work should monitor how this evolves during COP27 and onwards. Possible questions include (1) whether ideological minorities are growing or declining in influence, (2) whether social media polarisation is having a broader impact on public debates, {and (3) whether ideological echo-chambers are becoming more or less isolated as climate communication strategies develop.} 

Finally, it is a value judgement as to what constitutes a healthy plurality of views on social media, or unhealthy polarisation. Consensus should not be expected \cite{pearce2017beyond,machin2013negotiating}. However, tracking trends in polarisation over time is critical for understanding the political context for accelerated climate action and how political actions may impact public opinion.

\providecommand{\noopsort}[1]{}\providecommand{\singleletter}[1]{#1}%

\renewcommand\refname{Methods References}
\makeatletter
\apptocmd{\thebibliography}{\global\c@NAT@ctr 56\relax}{}{}
\makeatother

\section*{Methods}

\subsection*{Datasets}

\subsubsection*{Twitter}
Twitter data including tweets and user information was collected using the official Twitter API for academic research (\url{https://developer.twitter.com/en/docs/twitter-api}), using the search query ``cop2x'', $x \in \{0,\ldots,6\}$. For each COP, data was collected from {June} 1st in the year of the conference, to May 31st the following year, with the exception of COP26 where data was collected up to and including November 14th, 2021. Statistics for each COP are provided in Supplementary Table~5. Each dataset was downloaded between October and November 2021.

\subsubsection*{Politicians on Twitter}

{Twitter accounts associated with elected politicians were labelled using an existing dataset of political Twitter handles from 26 countries collected between September 2017 and February 2021. The dataset is freely available at} \href{http://twitterpoliticians.org/download}{TwitterPoliticians.org} {or on} \href{https://figshare.com/articles/dataset/The_Twitter_Parliamentarian_Database/10120685}{FigShare}.
{The dataset is discussed in detail in} \cite{van2020twitter}. {Note that select politicians elected in 2021 missing from the dataset were added manually to the database if they appeared as prominent influencers in the COP26 network (e.g., @laurenboebert). For practical purposes, an account is labelled as a politician even if that politician no longer holds elected office.}

\subsection*{Network construction}

The Twitter interaction network is constructed by taking the full corpus of tweets for each COP and focusing exclusively on retweets. Such an approach is typical in the Twitter analysis literature, where retweets are considered evidence of a user endorsing the message of the original poster; this is despite many Twitter users stating in their biography that retweets should not be understood as endorsements. This is in contrast to quote tweets or comments which are less likely to represent a clear endorsement of a tweet. After selecting all the retweets from the full Twitter dataset, we filter by language using the Twitter API language metadata, selecting only those retweets written in English.

From this set of English language retweets, a network is constructed by defining a node for each unique user in the dataset. This includes any user who authored an original English language tweet, or retweeted an English language tweet, containing the keyword ``cop2x'', $x \in \{0,\ldots,6\}$. A directed edge is formed from node A to node B if user A retweeted a post authored by user B. Edges are weighted according to the number of unique retweets between those two users.

\subsection*{Measuring polarisation}

{Assessments of polarisation in social systems have become a key research theme in computational social science, particularly in recent years following the Trump presidency} \cite{waller2021quantifying,flamino2021shifting} 
{in the USA and Brexit in the UK}  \cite{del2017mapping}. {Despite this, there is no one agreed definition of polarisation, with variable definitions depending on research question and field.}

{In the social sciences, the term polarisation is typically understood as some form of distance measure on a (typically one-dimensional) distribution of opinions. Under this general framework, polarisation may be quantified in numerous ways including, but not limited to: spread, dispersion, regionalisation, community fracturing, distinctness, and group size. For an extensive discussion of these ``senses'' of polarisation, see }\cite{bramson2017understanding}. {As stated in} \cite{bramson2017understanding}, {``the most common measure of polarisation in the political literature
is probably bimodality, which is the idea that the population can be usefully
broken down into two subpopulations''.} {This is the definition we choose in the current paper, in particular because it reflects how prominently politicians (for whom Twitter is particularly influential) see content which is pro-climate or climate sceptical. We stress, however, that using alternative definitions of polarisation may lead to our results being interpreted differently.} 

{One of the limitations in this family of polarisation measures is that they typically consider opinion distributions in the absence of structure; we may observe polarised views amongst individuals, but we do not necessarily know how different individuals interact with each other. It is this structural factor that is focused on in network science, where polarisation is often thought of as a distance measure on two (or more) network communities; for a nice example of such work with relevance to climate change see} \cite{chen2021polarization}. {However, this structural point of view (polarisation in terms of interactions) often fails to consider polarisation in terms of opinions. This is a limitation, but reflects the reality of most social media studies of polarisation where structure is known, but ground truth opinion (e.g., from surveying individuals) is not known. This is the case for Twitter where the ``true'' opinion of an account is unknown.}

{The latent ideology measure} \cite{barbera2015tweeting} {used in the current study aims to infer a synthetic opinion distribution from network structure, based on the premise that the structural separation of group interactions on a particular topic should correlate with differences in group opinions on that topic. Without external validation, such synthetic opinion distributions can be dangerous, particularly if a network appears structurally polarised for reasons other than individual views on a topic (for instance, due to geographical factors). However, with validation, such an approach has the benefit of combining the nuanced social science concept of polarisation with the structural approach typical in social media studies.}

{Having extracted the distribution of opinions using the latent ideology method, we quantify polarisation in terms of bimodality using Hartigan's diptest (see below). The choice to measure polarisation in terms of bimodality is deliberate since it is a relative measure (as opposed to an absolute measure) which can be applied to a synthethic distribution of opinions. Using absolute measures is difficult with synthetic distributions given that absolute opinion scores are not easily mapped to scores which may be derived from surveys (e.g., out of 10, how strongly do you support climate action?). }

\subsubsection*{Latent ideology}
The latent ideology estimation was developed in \cite{barbera2015birds,barbera2015tweeting} and adapted for exploiting retweet interactions in \cite{flamino2021shifting}. Following \cite{flamino2021shifting}, we infer ideological scores for Twitter users using correspondence analysis~\cite{benzecri1973analyse} (CA) and retweet interactions.

First, we build a matrix $\matr{A}$ such that each element $a_{ij}$ is the number of times user $i$ retweeted influencer $j$. To select only users that are interested in the COP26 debate, we prune out users that retweeted fewer than two influencers. 

We then execute the CA method according to the following steps.
Given the adjacency matrix normalised by the total number of retweets as $\matr{P}=\matr{A}(\sum_{ij}a_{ij})^{-1}$, the vector of row and column sums respectively as $\matr{r}=\matr{P}\matr{1}$ and $\matr{c}=\matr{1}^T\matr{P}$, and considering the matrices $\matr{D}_r=\text{diag}(\matr{r})$ and $\matr{D}_c=\text{diag}(\matr{c})$, we can compute the matrix of standardised residuals of the adjacency matrix as $\matr{S}=\matr{D}^{-1/2}_r(\matr{P}-\matr{r}\matr{c})\matr{D}^{-1/2}_c$.
The usage of the standardised residual matrix allows the method to account for differences in users' activity and influencers' popularity.
Next, single value decomposition is applied to the matrix $\matr{S}$ as
$\matr{S}=\matr{U}\matr{D}_{\alpha}\matr{V}^T$ with $\matr{U}\matr{U}^T=\matr{V}\matr{V}^T=\matr{I}$ and $\matr{D}_\alpha$ being the singular values diagonal matrix.
The standard row coordinates $\matr{X}=\matr{D}_r^{-1/2}\matr{U}$ can be considered as the estimates of the user ideologies. In our study, we only consider the first dimension that corresponds to the largest singular value. 
Users' ideological positions are computed by rescaling the row estimates into the set $[-1,1]$, while the influencers' ideological positions are calculated by the median of the weighted position of their retweeters.

\subsubsection*{Hartigan's diptest}
Hartigan's diptest is a nonparametric test to measure the multimodality of a distribution from a sample \cite{hartigan1985dip}. It calculates the maximum difference over all sample points between the unimodal distribution function that minimises that maximum difference and the empirical distribution function. The test produces a statistic $D$ which quantifies the magnitude of multimodality, and a statistical significance $p$. If $p < 0.01$, we say that the ideology distribution shows statistically significant multimodality. Conversely, if $p \geq 0.01$, we cannot reject the unimodality of the distribution. 

{The diptest calculates the test statistic, $D$, from the full set of influencer and user ideology scores. In order to estimate errors for the diptest we use a bootstrapping procedure. This involves selecting 70\% of the users and influencers at random from the pre-computed ideology scores, and recalculating the diptest from this sample. Repeating the sampling process one thousand times gives a distribution of diptest scores from which diptest errors can be computed.}

\subsubsection*{Selecting influencers}

Applying the latent ideology to a set of influential accounts on Twitter does not guarantee that those accounts will arrange themselves in the latent space based on political or climate ideology. In a number of cases, the dominant factor which determines the principal ideological axis is geography. By focusing exclusively on English language Twitter, the effect of these geographic factors is reduced. However, some additional filtering is required to avoid the latent ideology partitioning accounts based on geography.

Factors which may conflate ideological scores include (1) language (e.g., English vs. non-English), (2) geography (e.g., accounts focused on Indian politics), and (3) prominent topics outside the core discussion (e.g., discussions in the blockchain community), see SI section 2B for details. These factors are mitigated by selecting English language tweets, and by performing some minor filtering of the influencer set. For each COP, less than $3\%$ of accounts are removed from the set of influencers as part of the filtering process.

{Note, in the SI section 1C we discuss other influencer definitions and show that the observed increase in polarisation during COP26, relative to previous COPs, is robust across a range of measures (Supplementary Figures 3--7).}

\subsection*{Topic extraction using BERT}

{BERTopic} \cite{grootendorst2022bertopic} {is a novel topic modelling tool that extracts latent topics from a collections of documents. The base algorithm uses pre-trained transformer-based language models to build document embeddings, and produces topic representations by clustering embeddings and applying a class-based TF-IDF procedure} \cite{claude2010tfidf}. 

{BERTopic is well-suited to analysing Twitter data where tweets naturally act as documents such that coherent and consistent themes can be derived from the text due to its ability to generate sentence vector representations which can preserve semantic structure. In contrast, traditional topic modelling typically uses the bag-of-words approach to define topics based on word frequency.}

\subsection*{News media URL classification}
To highlight the different news sources used by the ideological minority and majority, we exploited data retrieved from NewsGuard \url{https://www.NewsGuardtech.com/}.
NewsGuard is a tool that provides trust ratings for news and information websites. NewsGuard assesses the credibility and transparency of news and information websites based on nine journalistic criteria. These criteria are individually assessed and then combined to produce a single ``trust score'' from 0 to 100 for a given news media outlet. Scores are assigned by a team of journalists, not algorithmically. Scores are not given to platforms (e.g., Twitter, Facebook), individuals, or satire content. More detail regarding the rating process is available at \url{https://www.NewsGuardtech.com/ratings/rating-process-criteria/}. To complement news media trust scores, NewsGuard also provide a political leaning for news outlets (far left, slightly left, slightly right, far right) which allows us to gauge the ideological leaning of the news sources referenced in the COP Twitter discussion. Note that NewsGuard classifies a far larger set of news sources as slightly left of right, than far left or right.

Using the database of news media trust scores, we cross reference the domains found in individual tweets with the corresponding trust score from the NewsGuard database.
{For COP21, we have 5.7 million tweets (including non-English tweets), of which 3.8 million contain a URL. Of these URLs, we are able to classify 730 thousand using the NewsGuard dataset (19\% of tweets with a URL). In contrast, for COP26 we have 10.2 million tweets (including non-English tweets) of which 2.8 million contain a URL (note that far fewer tweets contain URLs relative to COP26). Of these 2.8 million URLs, we are able to classify 560 thousand (20\% of tweets with a URL) using the NewsGuard dataset.}

\section*{Extended data}

\renewcommand{\figurename}{Extended Data}
\setcounter{figure}{0}

\begin{figure}[h!]
    \centering
    \includegraphics[width=\linewidth]{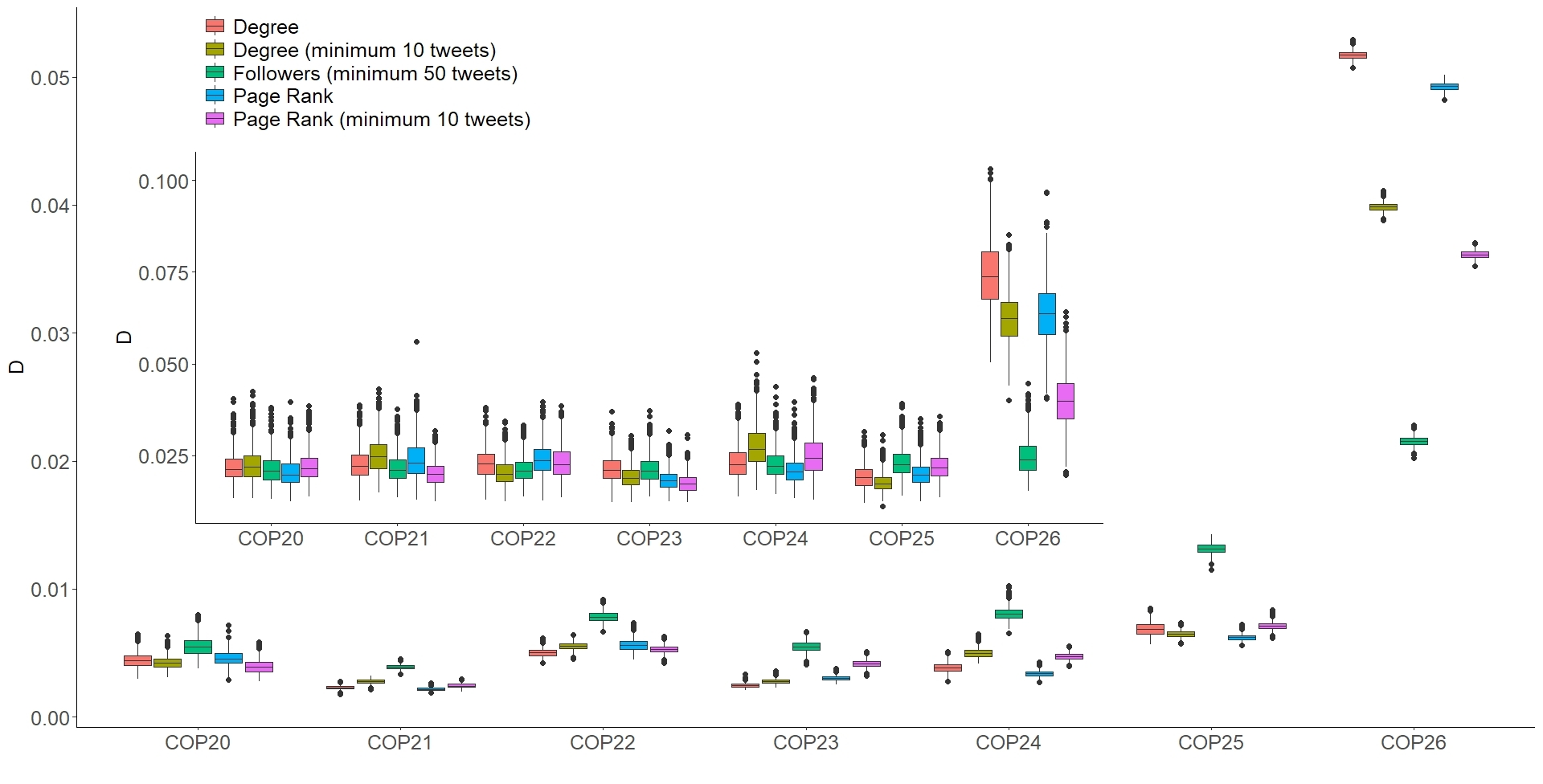}
    \caption{Hartigan's diptest statistic, $D$, from the latent ideology user distribution for each COP from COP20 -- COP26, calculated using five different influencer rankings. Inset: the equivalent for the influencer ideology. Note that using the followers ranking, most minority influencers are lost in COP26, with only minority news agencies and select politicians remaining. However, user ideology which measures polarisation across the network as a whole continues to show an increase in polarisation during COP26, relative to previous COPs. Note, the degree is equivalent to the number of retweets. For a full discussion of this figure see SI section 1C.}
    \label{figsm:influencer_robust}
\end{figure}

\begin{figure}
    \centering
    \includegraphics[width=\linewidth]{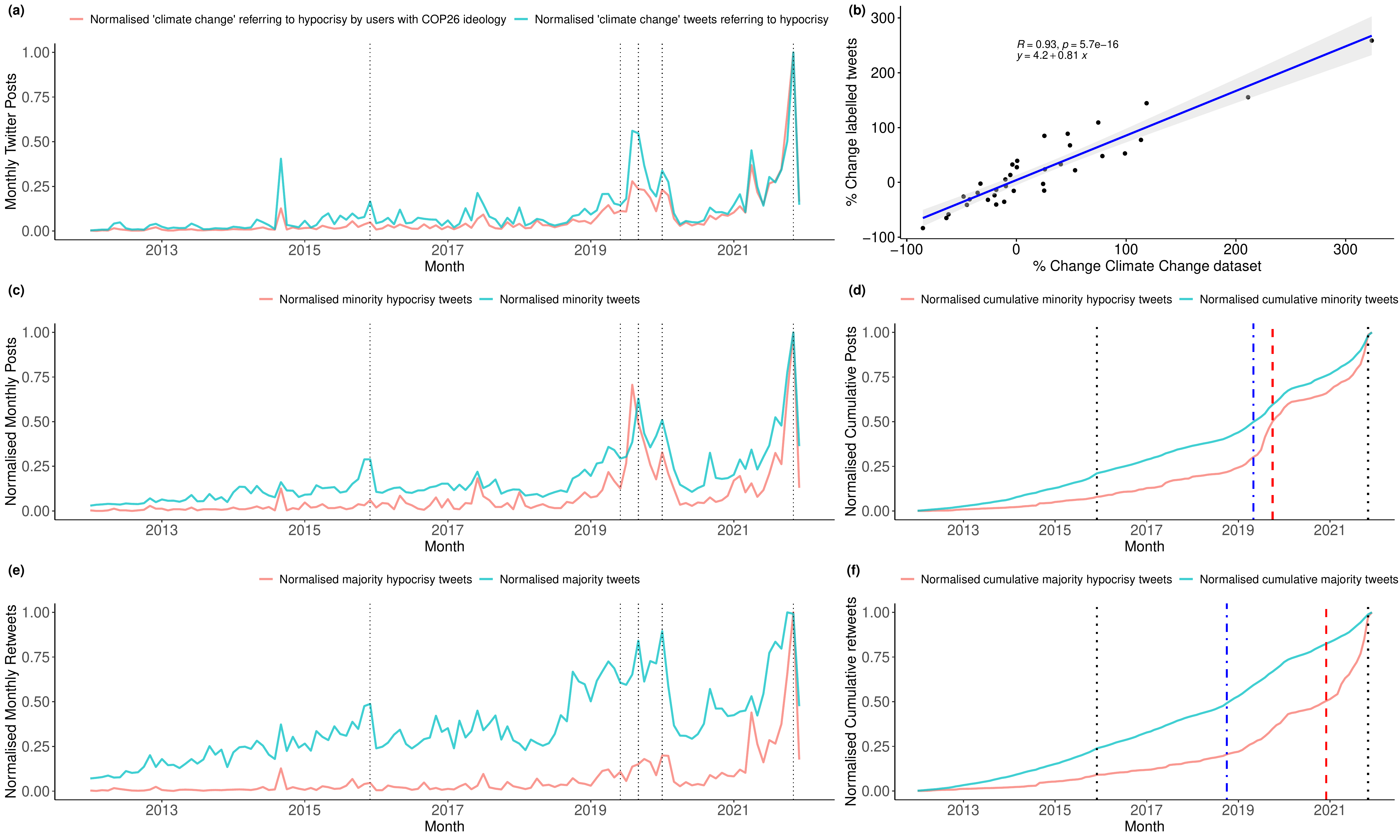}
    \caption{\textcolor{black}{\textbf{Tweets in the ``climate change'' dataset referring to hypocrisy or related terms.} (a) The normalised number of tweets each month referring to themes of hypocrisy for all users in the climate change dataset in blue, and for users with a COP26 ideology score in red. (b) The monthly percentage change for each of the curves in panel (a) showing that the activity of users with a labelled COP26 ideology is highly correlated with general activity, Pearson's $R = 0.93$. Note that due to very low tweets counts referring to hypocrisy in some months ($<10$), Pearson's R is calculated using data from 2019 onwards. 
    (c) Tweets from the COP26 minority in the ``climate change'' dataset are shown in blue, and for tweets which refer specifically to themes related to hypocrisy are shown in red.
    (d) Normalised cumulative curves for the data in panel (c). 
    (e) Tweets from the COP26 majority in the ``climate change'' dataset in blue, and for tweets which refer specifically to themes related to hypocrisy in red.
    (f) Normalised cumulative curves for the data in panel (e). 
    Black dotted lines on the left correspond to key events: (1) COP21 in December 2015, (2) the Canada oil pipeline announcement in June 2019, (3) the global climate protests in September 2019, (4) the Australian bushfires in January 2020, and (5) COP26. The black dashed lines on the right indicate COP21 and COP26. The red and blue dashed lines mark the median of the cumulative curves for the majority and minority respectively. For a full discussion of this figure see SI section 1K.}}
    \label{figsm:hypocrisy}
\end{figure}

\begin{figure}
    \centering
    \includegraphics[width=\linewidth]{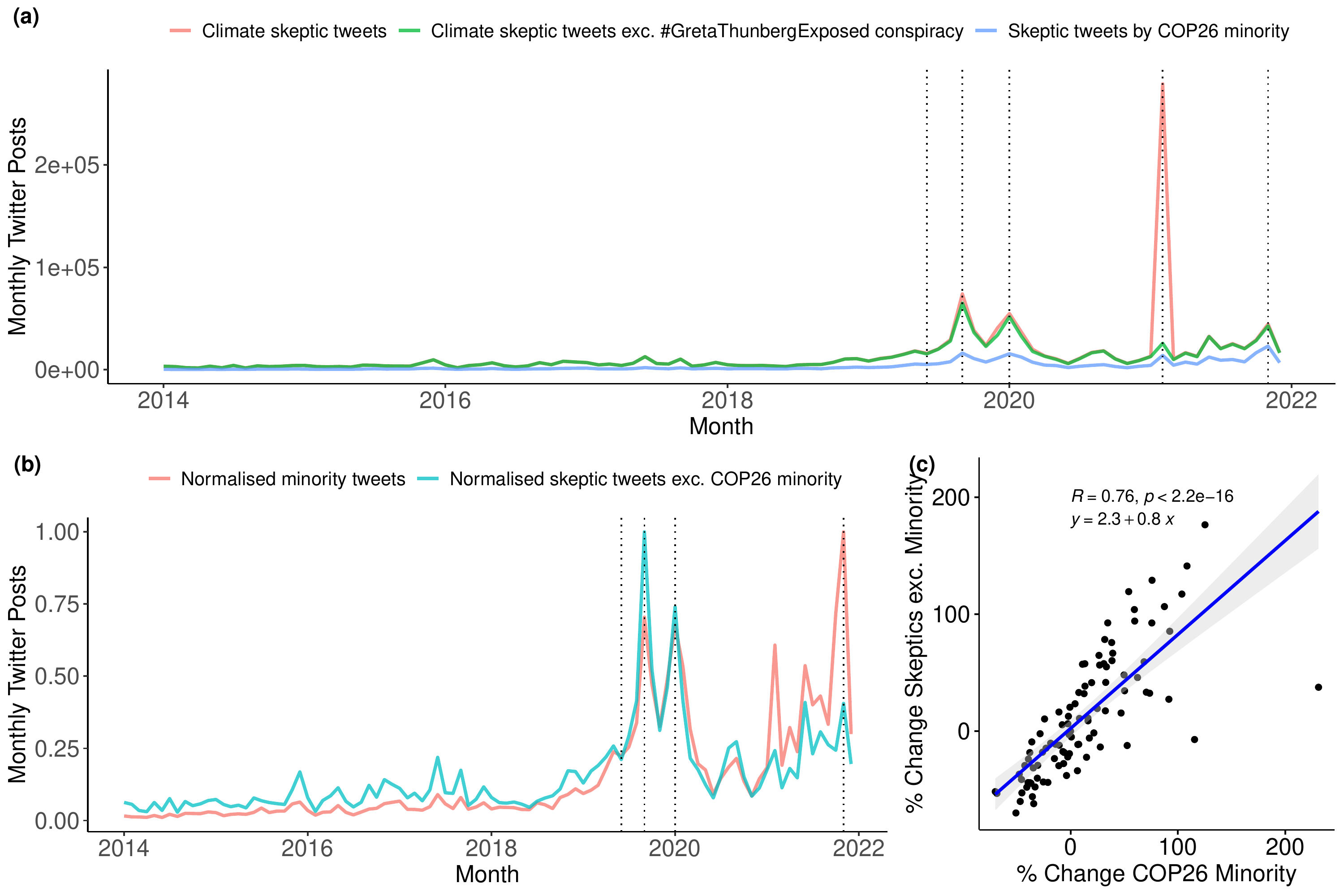}
    \caption{\textcolor{black}{\textbf{Tweets using terms associated with climate scepticism.} (a) The total number of climate sceptic tweets each month in red, and after filtering out the \#GretaThunbergExposed conspiracy theory that was specific to India in February 2021 in green. Monthly tweets by users from the COP26 minority are shown in blue. (b) The normalised number of climate sceptic tweets by users from the COP26 minority in red, and by all other users, excluding India, in blue. (c) The percentage change in the time series shown in panel (b), fitted using linear regression, Pearson's $R = 0.76$, showing that climate sceptics in the COP26 minority are highly correlated with those who are not part of the COP26 minority. This proves that the COP26 minority are representative of climate scepticism on Twitter in general. The dashed lines correspond to events of particular importance: (1) Justin Trudeau's announcement of an oil pipeline in June 2019, (2) the global climate protests in September 2019, (3) the Australian bushfires, (4) the \#GretaThunbergExposed conspiracy in India in February 2021, and (5) COP26. For a full discussion of this figure see SI section 1K.}}
    \label{figsm:skeptics}
\end{figure}

\begin{figure}
    \centering
    \includegraphics[width=\linewidth]{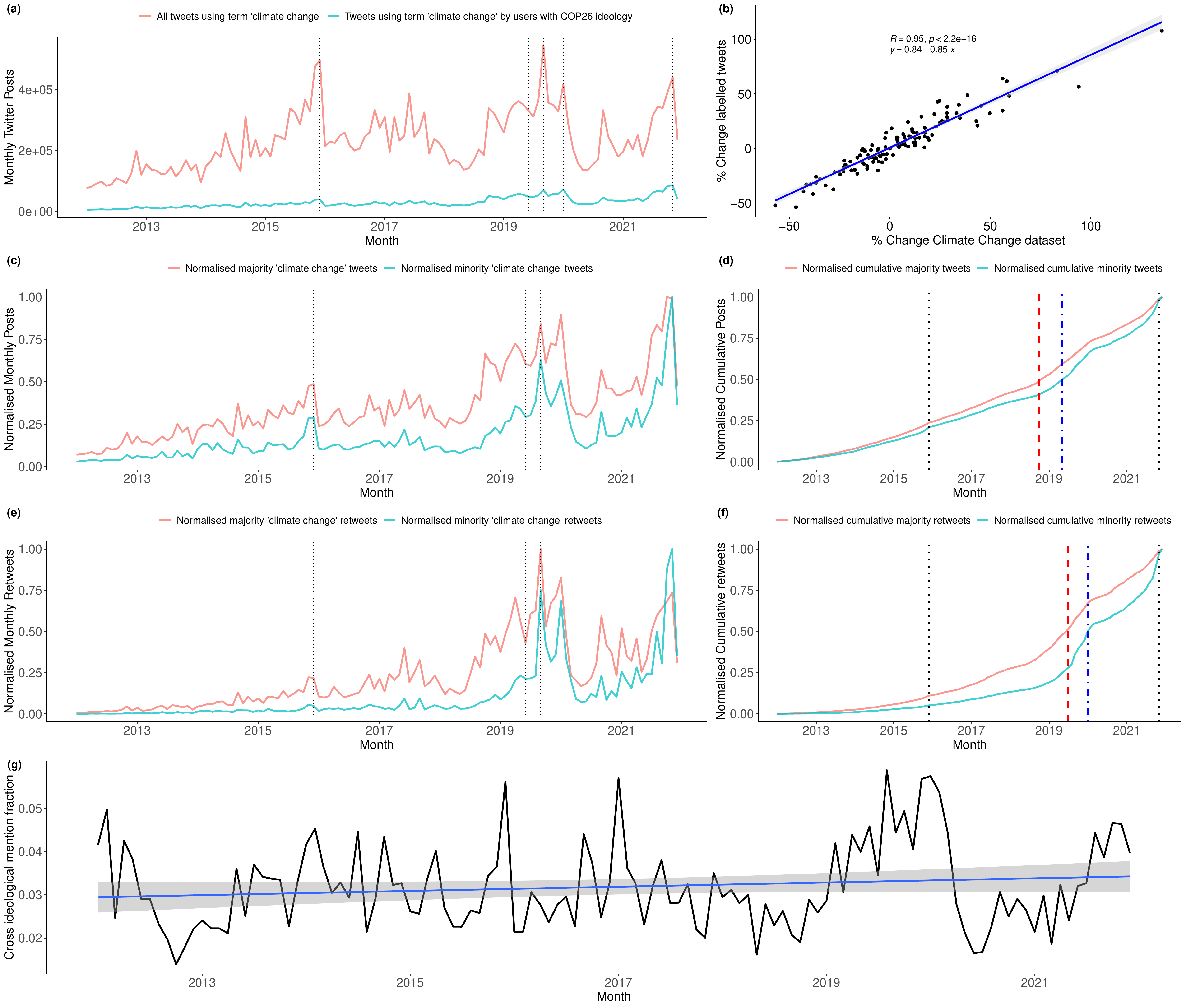}
    \caption{\textcolor{black}{\textbf{Tweets using the term ``climate change''.} (a) The total number of tweets each month since 2010 using the term ``climate change'', and tweets by users with a COP26 ideology score. (b) The monthly percentage change for each of the curves in panel (a) showing that the activity of users with a labelled COP26 ideology is highly correlated with general activity, Pearson's $R = 0.94$. (c) Tweets using the term ``climate change'' by users with a majority (red) and minority (blue) ideology score. (d) Normalised cumulative curves for the data in panel (c). (e) The number of retweets received by tweets using the term ``climate change'' by users with a majority (red) and minority (blue) ideology score. (f) Normalised cumulative curves for the data in panel (e). (g) The fraction of user mentions by members of the COP26 minority and majority which are cross-ideological, fitted using ordinary least squares. 95\% confidence interval on gradient: [-3.6e-5, 3.0e-5] percent per day, confirming that there is no evidence of an increase in engagement between majority and minority groups over time. Black dotted lines on the left correspond to key events: (1) COP21 in December 2015, (2) the Canada oil pipeline announcement in June 2019, (3) the global climate protests in September 2019, (4) the Australian bushfires in January 2020, and (5) COP26. The black dashed lines on the right indicate COP21 and COP26. The red and blue dashed lines mark the median of the cumulative curves for the majority and minority respectively. For a full discussion of this figure see SI section 1K.}}
    \label{figsm:climate}
\end{figure}

\clearpage

\section*{Data availability}
Expanded figures for the latent ideology are available for each COP at \cite{falkenberg2021growingDATA}. Twitter and Youtube data is made available in accordance with Twitter and Youtube's terms of service. Tweet and Youtube video IDs for each COP are available at \cite{falkenberg2021growingDATA}. The corresponding tweets can be downloaded using the official Twitter API (\url{https://developer.twitter.com/en/docs/twitter-api}). Youtube video metadata can be downloaded using Youtube's official API (\url{https://developers.google.com/youtube/v3}). Reddit data was downloaded using the \url{https://pushshift.io/} API, and is freely available to the public. 

\section*{Code availability} 
The \texttt{R} code used to calculate the latent ideology is available at \cite{falkenberg2021growingDATA}. However, due to Twitter's Terms of Service, we are unable to provide the retweet networks required as an input to the latent ideology code. Tweets can be downloaded using Twitter's API using the IDs provided (see data availability). The OSF repository includes a dummy retweet network and influencer list to illustrate the required data format for the latent ideology code.

\section*{Acknowledgements}
M.F., A.G., M.T., F.Z., W.Q. and A.B. are grateful for support from the IRIS Infodemic Coalition (UK Government, no. SCH-00001-3391). F.Z. acknowledges financial support from the European Union’s Rights, Equality and Citizenship project EUMEPLAT grant no. 101004488.

\section*{Author contributions statement}

M.F., A.G., F.Z., W.Q. and A.B. conceptualised the paper and designed the experiments. M.F. and A.G. acquired the data. M.F., A.G. and M.T. carried out the experiments. M.F., A.G., M.T., F.L., W.P., F.Z., W.Q. and A.B. contributed to the interpretation of results. M.F., A.G., M.T., N.D.M. and M.S. implemented the data presentation and visualisation. M.F., A.G., M.T., N.D.M., F.L., M.S. and A.M. contributed to the supplementary information. M.F., A.G., W.P., F.Z., W.Q. and A.B. wrote the main manuscript. F.Z., W.Q. and A.B. acquired funding for the project.

\section*{Competing interests statement}
The authors declare no competing interests.

%\bibliography{apssamp}


\providecommand{\noopsort}[1]{}\providecommand{\singleletter}[1]{#1}%

\end{document}